\pdfoutput=1
\documentclass{raa}          

\usepackage{graphicx,times}             
\usepackage{natbib}
\usepackage{amssymb,amsmath}
\usepackage{float}
\usepackage{bbm}
\usepackage{latexsym}
\bibpunct{(}{)}{;}{a}{}{,}

\usepackage[colorlinks,urlcolor=blue,citecolor=blue,pagebackref=true]{hyperref}

\begin{document}

  \title{Transverse Velocity Field Measurement in High-Resolution
         Solar Images Based on Deep Learning
}

   \volnopage{Vol.0 (20xx) No.0, 000--000}      
   \setcounter{page}{1}          

   \author{Zhen-Hong Shang 
      \inst{1,2,}\footnote[1]{Corresponding author}
   \and Si-Yu Mu 
      \inst{1}
   \and Kai-Fan Ji 
      \inst{3}
   \and Zhen-Ping Qiang 
      \inst{4}
   }

   \institute{Faculty of Information Engineering and Automation, Kunming University of Science and Technology, Kunming, 650500, China; {\it szh@kust.edu.cn}\\
        \and
            Yunnan Key Laboratory of Artificial Intelligence, Kunming University of Science and Technology, Kunming, 650500, China\\
        \and
            Yunnan Observatories, Chinese Academy of Sciences, Kunming 650216, China\\
        \and
            College of Big Data and Intelligent Engineering, Southwest Forestry University, Kunming 650224, China\\
\vs\no
   {\small Received 20xx month day; accepted 20xx month day}}

\abstract{ To address the problem of the low accuracy of transverse velocity field measurements for small targets in high-resolution solar images, we proposed a novel velocity field measurement method for high-resolution solar images based on PWCNet. This method transforms the transverse velocity field measurements into an optical flow field prediction problem. We evaluated the performance of the proposed method using the H$\alpha$ and TiO datasets obtained from New Vacuum Solar Telescope (NVST) observations. The experimental results show that our method effectively predicts the optical flow of small targets in images compared with several typical machine- and deep-learning methods. On the H$\alpha$ dataset, the proposed method improves the image structure similarity from 0.9182 to 0.9587 and reduces the mean of residuals from 24.9931 to 15.2818; on the TiO dataset, the proposed method improves the image structure similarity from 0.9289 to 0.9628 and reduces the mean of residuals from 25.9908 to 17.0194. The optical flow predicted using the proposed method can provide accurate data for the atmospheric motion information of solar images. The code implementing the proposed method is available on \url{https://github.com/lygmsy123/transverse-velocity-field-measurement}.
}

   \authorrunning{Z.-H. Shang, S.-Y. Mu, K.-H Ji \& Z.-P. Qiang }            
   \titlerunning{Transverse Velocity Field Measurement }  

   \maketitle

%
%
\section{INTRODUCTION}           
\label{sect:intro}
The advent of high-resolution solar images captured by multi-band telescopes has revolutionized the study of the solar atmosphere, affording researchers an unprecedented level of detail and insight into the physical properties of the solar magnetic field. This wealth of data is the result of advanced imaging processing technology and equipment, which has enabled the study of the solar atmosphere to enter a new era of high temporal and spatial resolution. The measurement of the transverse velocity field of these high-resolution images has become a widely utilized method for analyzing the dynamics of the photosphere and chromosphere. 
For example, \citet{Schlichenmaier+Schmidt+2000} combined transverse and line-of-sight velocities to reconstruct the magnitude and orientation of the penumbral flow field in the deep photosphere, which is a significant step towards an understanding of the mass balance of the Evershed flow. \citet{Ruan+etal+2014} modeled the rotational motion of sunspots and the helical motion of solar filaments by transverse velocities, demonstrating that the rotation of the sunspots plays an important role in twisting, energizing and destabilizing the coronal filament-flux rope system during the solar eruption event. \citet{Yan+etal+2015} used the transverse velocity field to analyze the horizontal motion of the solar surface and found that the shearing motion of the opposite magnetic polarities and the rotation of the sunspots play an important role in the formation of filaments in the active region. High spatial and temporal resolution images place higher demands on the accuracy and efficiency of image transverse velocity field measurements.

The methods used to measure the transverse velocity field can be divided into traditional methods and machine-learning-based methods. Traditional methods have undergone significant development in the field of solar image processing. Traditional methods, such as local correlation tracking (LCT) (\citealt{November+Simon+1988}) and Fourier local correlation tracking (FLCT) (\citealt{Fisher+Welsch+2008}), use correlation functions to determine the maximum correlation position in the local region. Differential affine velocity estimation (DAVE) (\citealt{Schuck+2005}) is another method that uses an equation that can be directly solved by standard or total least-squares equations to calculate the velocity field. While these methods can measure large displacements and differences in images, they are not well suited for high-resolution solar images, which are characterized by strong self-similarity and many small motions, leading to inaccurate measurements of the velocity field.

Some traditional methods transform the computation of the transverse velocity fields to that of an optical flow field calculation. Optical flow refers to the instantaneous velocity of pixel motion in the imaging plane of a time-series image and is therefore considered a two-dimensional immediate velocity field composed of all pixel points in the image. Traditional optical flow methods follow the principles of the pioneering work of 
\citet{Horn+Schunk+1981} 
and calculated the optical flow based on various assumptions. Some studies have used a coarse-to-fine framework that relies on energy minimization methods to estimate the optical flow 
(\citealt{Brox+etal+2004, Brox+Malik+2010}). 
Some studies have introduced descriptor-matching techniques 
(\cite{Steinbrucker+etal+2009, Liu+etal+2010}), 
to obtain correspondence by computing the correlation between descriptors. However, the coarse-to-fine framework leads to the loss of small targets and prevents the accurate prediction of optical flow. Moreover, descriptor-matching methods fail to deal with the mismatching problem when facing dense optical flow prediction, resulting in accuracy loss.
 
The Demons algorithm is used in this work to create optical flow datasets. (see the \nameref{sect:data} section). The major principles of the algorithm are provided below for the sake of completeness. Demons (\citealt{Thirion+1998}) treats the optical flow field as forces that drive pixel movements. It uses the pixel differences $I_1-I_2$ between two consecutive frames as the external forces and the gradient information of the image $\nabla I$ as the internal forces. The dense optical flow is generated by iteratively calculating the displacement of each pixel as follows:
\begin{equation}\label{eq1}
	\mathbf{v}=\left(I_1-I_2\right)\cdot
	\left[\frac{\nabla I_1}{\|\nabla I_1 \|^2 + \alpha^2\left(I_1-I_2\right)^2} + 
	\frac{\nabla I_2}{\|\nabla I_2 \|^2 + \alpha^2\left(I_1-I_2\right)^2}\right],
\end{equation}
where $\alpha$ is the normalization factor, which controls the magnitude of the driving force. However, the pixels in $I_1$ can move freely, which may cause pixels with the same intensity value to be mapped to the same pixel point in $I_2$, resulting in incorrect matching. To address this issue, \citet{Liu+etal+2018} utilized gradient mutual information to further improve the Demons method enabling image pixels to move in the correct direction. The improved Demons algorithm is formulated as follows:
\begin{equation}\label{eq2}
	\begin{aligned}
	\mathbf{v}_{n+1}&=G_{\sigma}\bigotimes \Bigg\{ \mathbf{v}_n + \left(I_1-I_2\right)\cdot \left[\frac{\nabla I_1}{\|\nabla I_1 \|^2 + \alpha^2\left(I_1-I_2\right)^2}
	+ \frac{\nabla I_2}{\|\nabla I_2 \|^2 + \alpha^2\left(I_1-I_2\right)^2}\right]\\
	&+ \beta\max \left[I_{MI}\left(\mathbf{v}_{n}\right)\right] \Bigg\},
	\end{aligned}
\end{equation}
where $\max \left[I_{MI}(\mathbf{v}_n)\right]$ represents the gradient mutual information of the two images, $\beta$ is a positive constant indicating the weight of this item, and $G_{\sigma}$ denotes Gaussian operation. The formula indicates the processing steps of the iterative calculation, and $\mathbf{v}_n$ is the optical flow field calculated by the nth iteration. However, the accuracy of the algorithm can still be improved when dealing with the non-rigid deformation in H$\alpha$ images.

With the remarkable success of the neural network AlexNet
(\citealt{Krizhevsky+etal+2017}) in image-classification tasks, deep learning has been studied and applied in many fields, thereby demonstrating its powerful performance. Some approaches use fully convolutional neural networks to predict optical flow. 
\citet{Dosovitskiy+etal+2015} proposed FlowNet, which pioneered the use of a U-Net encoder--decoder structure to predict the optical flow of two images. 
\citet{Ranjan+Black+2017} proposed SpyNet, which uses an image warping technique at each layer of the network by introducing an image pyramid structure to reduce the difference between two images frames to construct a lightweight optical flow-prediction network. However, FlowNet and SpyNet still face challenges in accurately predicting optical flow for small targets.

Several studies have contributed to the improvement of SpyNet prediction accuracy for optical flows. \citet{Sun+etal+2018} proposed PWCNet, which utilized local matching costs, feature warping, and cascade processing to obtain the residual flow. \citet{Hui+etal+2018} proposed LiteFlowNet, which used feature warping and introduced flow field regularization into the network. However, PWCNet and LiteFlowNet primarily deal with large-displacement optical flow-prediction problems, and the optical flow field predicted at the edges of moving objects is too smooth to predict the optical flow of small targets, resulting in a loss of accuracy.

\citet{Teed+Deng+2020} proposed the Recurrent All-Pairs Field Transforms (RAFT), which introduces the idea of recurrent neural network (RNN) on convolutional neural networks (CNNs). RAFT iteratively refines the optical flow using a gate recurrent unit (GRU). However, the cost volume can only be constructed at a low resolution, causing a loss of feature information for tiny structures. The process of upsampling from low-resolution optical flow to full resolution also increases errors and decreases the accuracy of small target optical flow predictions.

Typical transverse-motion velocity values of the solar photosphere and chromosphere range from a few kilometers to tens of kilometers per second. This implies that we must accurately measure the displacement between two frames at a subpixel level in a sequence of high-resolution solar images. The high-resolution solar images exhibit strong self-similarity and contain numerous tiny structures, which existing deep learning-based optical flow methods struggle to predict accurately. In response to this, we propose a novel optical flow prediction method based on PWCNet to address the issue of low accuracy for small target predictions in high-resolution solar images.

The remainder of this paper is organized as follows. Section 2 describes the data used in this study. In Section 3, we describe the proposed network. The experimental environment, results, and analysis of the proposed method are presented in Section 4. Finally, the conclusions are presented in Section 5.


\section{DATA}
\label{sect:data}
   \begin{figure}[h]
   \setlength{\abovecaptionskip}{0.cm}
   \setlength{\belowcaptionskip}{-0.cm}
   \centering
   \includegraphics[width=40em, angle=0]{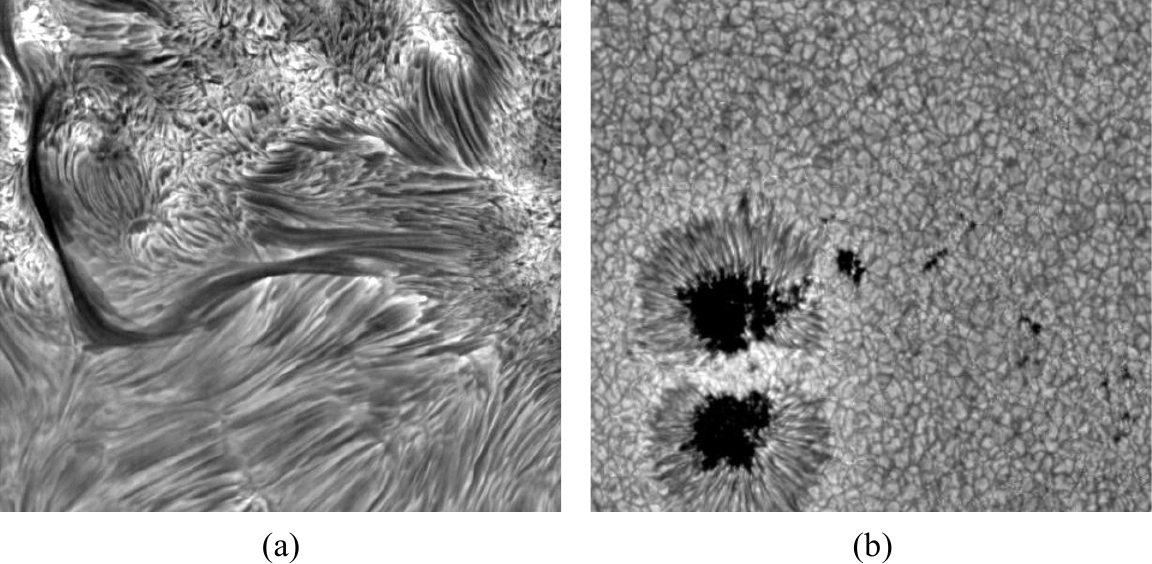}
   
   \caption{Second frame of Dataset 1 (a) and the second frame of Dataset 2 (b). }
   \label{Fig1}
   \end{figure}

The NVST is the largest solar telescope in China and one of the three largest high-resolution imaging solar telescopes worldwide. Their excellent observational performance and high-quality images have been confirmed in many respects. The data used in this study were obtained from the NVST at the Fuxian Solar Observatory of the Yunnan Astronomical Observatory, CAS. Two representative datasets of NVST observations were selected for this study, including photospheric and chromospheric images. The corresponding velocity field was obtained by calculating the optical flow field between the two image frames.

Dataset 1 contains Level 1+ reconstructed images of the chromospheric observed by NVST in the H$\alpha$ band from 07:26:20 on October 3, 2014 (UT) to 09:06:04 on October 3, 2014 (UT), with a time interval of 12 s for a total of 492 images. The field of view was 550×600 pixels, with an image resolution of 0.13 arcsec. The transverse velocity of 5 km/s on the chromosphere brings about a displacement of 0.6 pixels, while there is a filament eruption in this image sequence, and the overall optical flow field includes subpixel and superpixel displacements.

Dataset 2 consisted of Level 1+ reconstructed images of the photosphere observed by NVST in the TiO band from 01:28:04 (UT) to 04:32:07 (UT) on May 29, 2022, with a time interval of 36 s and a total of 301 images. The field of view was 404×468 pixels, with an image resolution of 0.052 arcsec. A typical transverse velocity of 1 km/s on the photosphere would bring about a displacement of 0.9 pixels, and the overall optical flow field includes subpixel and superpixel displacements.

Fig.~\ref{Fig1} illustrates a sample of the solar images used in this study. Fig. 1(a) shows a local region of the solar H$\alpha$ image with a strong self-similarity in the feature structure covering the filaments and fibers. Fig. 1(b) shows the local active region of the solar TiO image, and the characteristic structure covers the umbra and penumbra of the sunspots and granules.

In this study, the optical flow field corresponding to each group of images was calculated using the Demons method (\citealt{Liu+etal+2018}), and predicted images were generated according to the calculated optical flow field. The original image sequence, corresponding predicted image sequence, and predicted optical flow field were used as the training and testing datasets to form a proxy dataset that guided the training of the network. To guarantee that the training and testing datasets have diverse image distributions, Dataset 1 was split into a 9:1 ratio for training and testing based on time order, with 442 image pairs for training and 49 image pairs for testing. We used 300 pairs of images from Dataset 2 as the test set to measure the generalization performance of our method. It is emphasized that to measure the proposed method's generalization performance, the optical flow field of Dataset 2 is calculated without retraining the model but directly using the trained model of Dataset 1.

\section{METHOD}
\label{sect:method}
   \begin{figure}[h]
   \setlength{\abovecaptionskip}{0.cm}
   \setlength{\belowcaptionskip}{-0.cm}
   \centering
   \includegraphics[width=40em, angle=0]{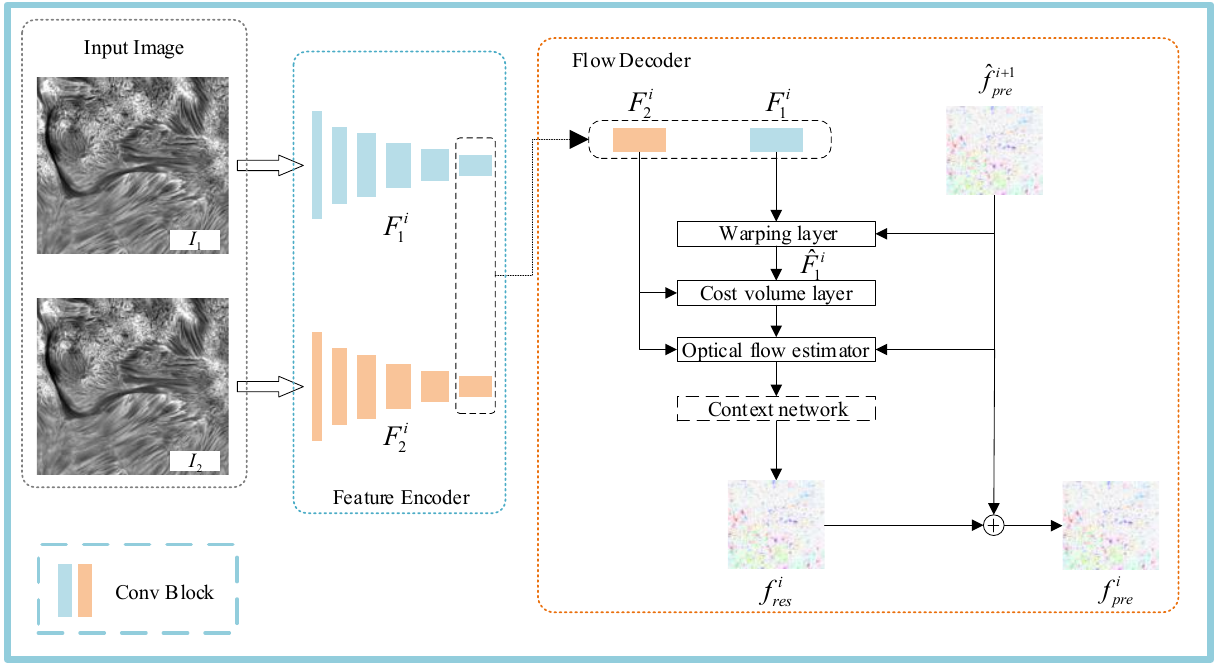}
   \caption{Structure diagram of PWCNet. }
   \label{Fig2}
   \end{figure}

   \begin{figure}[h]
   \setlength{\abovecaptionskip}{0.cm}
   \setlength{\belowcaptionskip}{-0.cm}
   \centering
   \includegraphics[width=40em, angle=0]{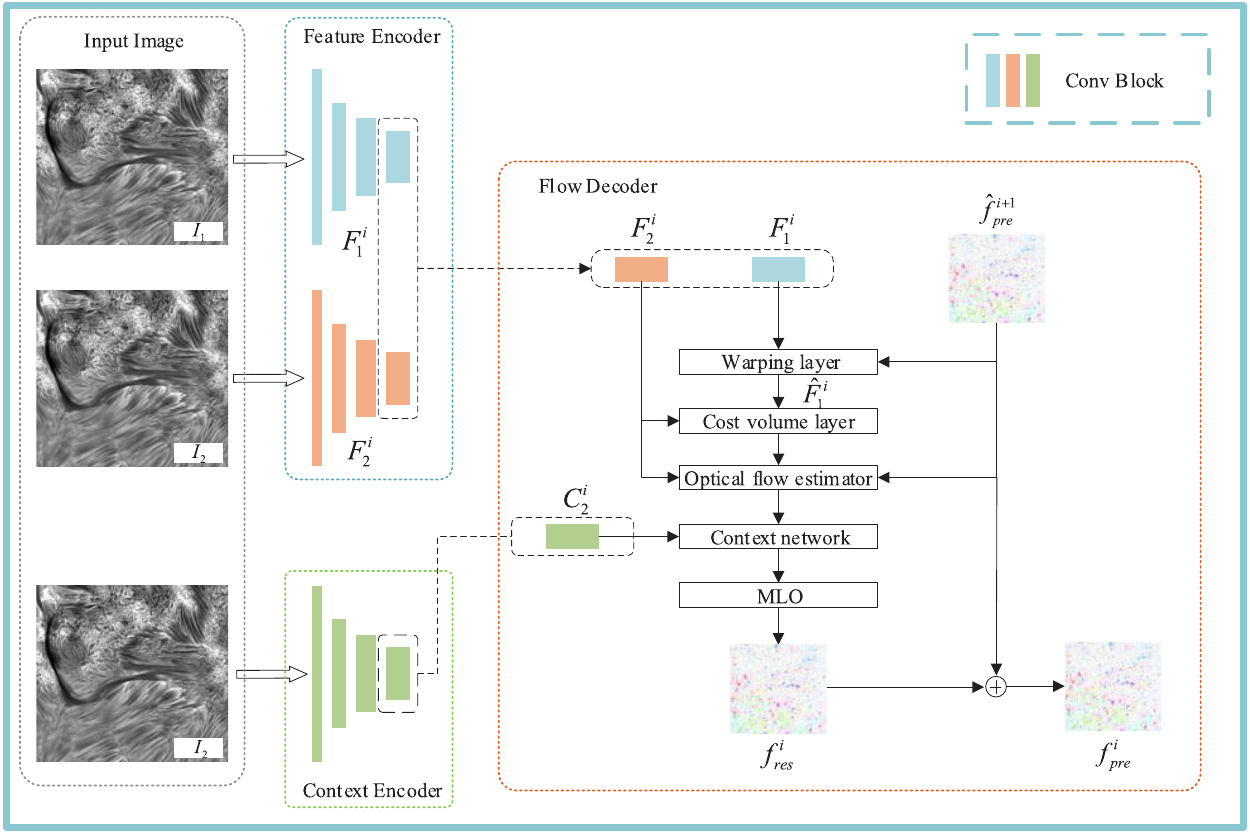}
   \caption{Structure diagram of improved network. }
   \label{Fig3}
   \end{figure}

\citet{Sun+etal+2018} proposed PWCNet, which has a pyramid structure and uses an encoder-decoder architecture, local matching cost, and feature warping techniques to estimate optical flow. The overall network architecture of PWCNet is illustrated in Fig.~\ref{Fig2}. The input to the model is two consecutive frames, $I_1$ and $I_2$, of size $H\times W\times1$, where H and W are the height and width of the image, respectively; and the output is a predicted optical flow of size $h\times w\times2$, where $h=H/4$ and $w=W/4$. In the encoding stage, PWCNet performs plain and stride convolutions to extract a six-layer feature pyramid with different resolutions. The decoder upsamples the predicted optical flow from the previous layer, warps the feature pyramid, and estimates the residual optical flow by comparing the warped feature pyramid from the first frame with which obtained from the second frame. Although PWC-Net outperforms the previous optical flow prediction model FlowNet, it still lacks the ability to capture low-level details such as edges and structural information, causing coarse and inaccurate predictions.

In this study, PWCNet was modified to solve the optical flow field prediction problem for small targets. The structure of the improved PWCNet is illustrated in Fig.~\ref{Fig3}. In the feature encoder module, we modified the number of feature pyramid layers of the original network to four, $F_n^i,i\in\{1,2,3,4\},n\in\{1,2\}$, where n represents the frame number and i denotes the pyramid layer number, corresponding to full, 1/2, 1/4, and 1/8 resolutions. The use of full-resolution features enable the network to capture small-target information and improve the capability of the network in small-target optical flow predictions. Subsequently, we replaced the plain convolution block with a residual convolution block, which reduced the model parameters, decreased the optimization difficulty, and made the model more lightweight. To learn small-target information, we constructed context feature pyramid $C_2^i$ based on the feature encoder to enhance the ability of the network to predict small-target optical flow. In the decoder module, we concatenated the matching information with contextual features and subsequently calculated the convolution. Multi-output module with a dilation convolution structure was connected and followed the context network to enhance the stability of the network. The model eventually produced four scales of predicted flows with different assigned weights. In contrast to PWCNet, we assigned large weights to high-resolution optical flows and small weights to low-resolution optical flows. This weight-assignment scheme enables the network to learn coarse-scale optical flows early in training without affecting the subsequent learning of finer optical flows.

\subsection{Improvement of Feature Encoder}
\begin{figure}[H]
   \centering
   \includegraphics[width=25em, angle=0]{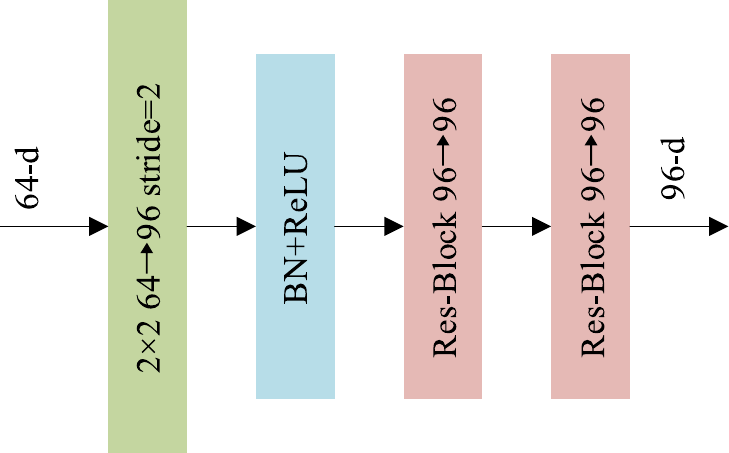}
   \caption{Structure diagram of feature encoder. }
   \label{Fig4}
   \end{figure}

The feature encoder of PWCNet network used six convolution blocks with a stride of two to obtain a feature pyramid with six layers of different resolutions. However, the direct application of stride convolution to the input image for downsampling causes the network to lose a large number of low-level features, such as the structural information of edges. We reduced the six layers of stride convolution to three layers and added plain convolution before the first stride convolution to obtain full-resolution features. The structure of the feature-extraction block for each feature encoder layer is illustrated in Fig.~\ref{Fig4}. Each feature-extraction block contains three convolutional layers, except for the first layer of the feature pyramid, where one layer has a kernel size of $2\times2$ and a stride of two to reduce the feature map size and transform the feature dimension, whereas the other two layers are residual blocks with a stride of one to obtain the residual information. Finally, the feature encoder generates a four-layer feature pyramid containing full, 1/2, 1/4, and 1/8 resolutions. High-resolution features acquired by the feature encoder capture a large amount of critical low-level feature information. By using this low-level information, the network can process a large amount of small-target motion information in high-resolution solar images and accurately predict small-displacement optical flow, which significantly improves the overall optical flow-prediction accuracy. Performing a convolution calculation at high resolution incurs a tremendous computational cost, and for the balance of computational cost and accuracy, fewer convolution blocks are used at the high-resolution layer.

\begin{figure}[H]
   \centering
   \includegraphics[width=30em, angle=0]{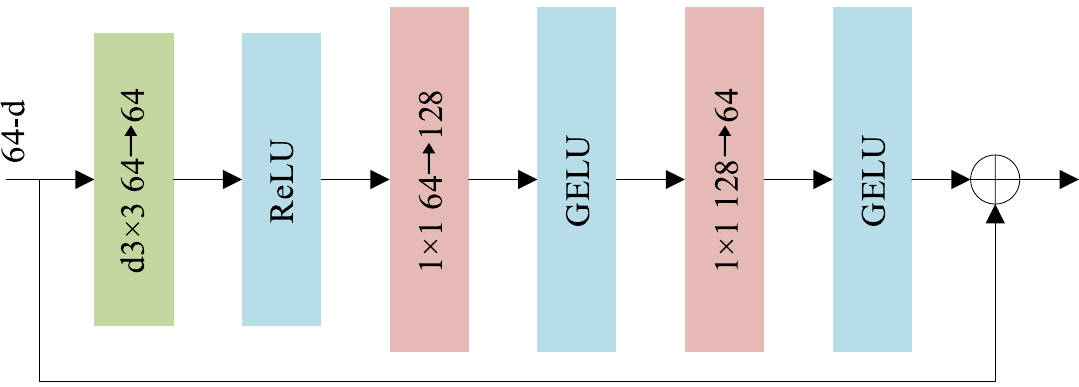}
   \caption{Structure diagram of residual block. }
   \label{Fig5}
   \end{figure}

The feature encoder of PWCNet was a stack of plain convolutional blocks. When the network becomes deeper, the gradients may disappear or explode, making parameter optimization difficult. For this problem, we used a residual convolution block to replace the plain convolution block and obtained accuracy gains from the increase in depth (\citealt{He+etal+2016}). The structure of residual block is illustrated in Fig.~\ref{Fig5}. In this structure, a $3\times3$ depthwise convolution layer is designed as the first layer of the residual block, and subsequently use two consecutive $1\times1$ plain convolution layers to fuse the channel features. This design further reduces the computational cost and number of network parameters, simplifies the optimization, and enhances network performance (\citealt{Liu+etal+2022}).

Furthermore, an context encoder was used to get the context feature of $I_2$ (\citealt{Teed+Deng+2020}), which allowed the network to obtain more small-target features, thus improving the accuracy of the optical flow prediction. The feature and context encoders together form the feature-extraction part of the framework.

\subsection{Improvement of Optical Flow Estimator}
The feed-forward connections in the optical flow estimation module and the cascaded dilated convolutions in the context module of PWCNet reduced the ability of the model to predict the optical flow. Therefore, it is important to consider enhancing the capability of the optical flow estimation module of the network when improving the module.

\begin{figure}[h]
   \centering
   \includegraphics[width=25em, angle=0]{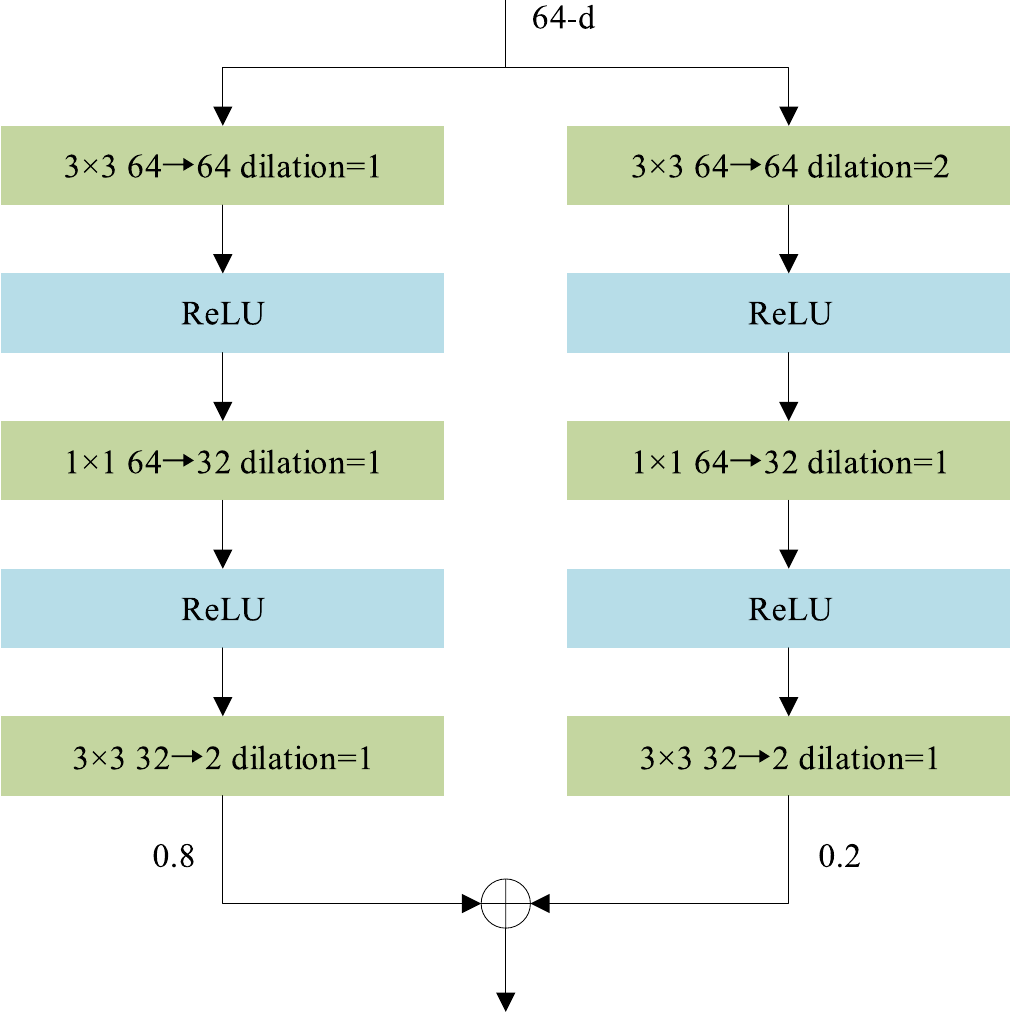}
   \caption{Structure diagram of multi-output module. }
   \label{Fig6}
   \end{figure}

To improve the expressive power of the network, there are two common approaches: increasing the depth or width of the network. The structure of the multi-output module is shown in Fig.~\ref{Fig6}. Multi-output module with different convolution kernels is introduced in the optical flow estimation module to obtain a larger receptive field and enhance the stability of the network (\citealt{Szegedy+etal+2015}). In the implementation, we applied a multi-output module with a convolutional kernel of $3\times3$, whose structure is illustrated in Fig.6. One branch utilizes a plain convolutional layer, and the other applies a dilated convolution with a dilation constant of two to enlarge the receptive field and enhance the expressive power of the network (\citealt{Hussain+etal+2022}). A $1\times1$ convolution was added following a $3\times3$ convolution to fuse the channel information and reduce the number of channels. In each branch, we output the corresponding residual optical flow. We assigned a large weight factor (0.8) to the optical flow of the plain convolution layer and a small weight factor (0.2) to that of the dilated convolution layer, according to the distance between the sampling point and the sampling location. Finally, we obtained this layer's residual optical flow ($f_{res}^{i}$) by summing the outputs of multiple branches.

\subsection{Improvement of Loss Function}
PWCNet uses the endpoint error (EPE) as the loss function to guide the training of the network, where EPE is the difference between the predicted optical flow $f_{pre}^i$ and the optical flow label $f_{gt}^i\in\mathbbm{R}^{H\times W\times2}$. The solar H$\alpha$ and TiO images exhibit strong self-similarity across structures, and the pixel motions remain relatively consistent within structures and vary widely between them. These characteristics made it difficult to rely solely on EPE to guide network training, which could result in the network's inability to learn the association of neighboring pixels and increase the optimization difficulty. 

Because the datasets used in our study are proxy datasets with some errors exist, additional loss functions are added to improve the model. In this study, we used three types of losses: endpoint error loss $\mathcal{L}_{epe}$, data item loss $\mathcal{L}_{data}$, and smoothness loss $\mathcal{L}_{smooth}$. The loss function $\mathcal{L}_i$ for layer i of the pyramid is formulated as follows:

\begin{equation}\label{eq3}
  \mathcal{L}_i = \omega_{epe}\mathcal{L}_{epe}^i + \omega_{data}\mathcal{L}_{data}^i + 
  \omega_{smooth}\mathcal{L}_{smooth}^i,
\end{equation}
where $\mathcal{L}_{epe}$, $\mathcal{L}_{data}$ and $\mathcal{L}_{smooth}$ are defined and explained in equations (\ref{eq4}), (\ref{eq5}) and (\ref{eq8}), respectively, and $\omega$ denotes the weight factor for each loss. In implementation, we set $\omega_{epe}$ to 1, $\omega_{data}$ to 0.05, and $\omega_{smooth}$ to 0.005. For the supervised network, $\mathcal{L}_{epe}$ avoids the manual setting of learning content and makes the model learn the essential correspondence of pixels; therefore, the $\omega_{epe}$ should be set to a larger value. $\mathcal{L}_{data}$ enables the predicted image $\hat{I}_1$ and target image $I_2$ to remain close in appearance, and $\mathcal{L}_{smooth}$ restricts the adjacent pixel motion to remain consistent. However, because of the existence of intensity variations in the data and to prevent over-smoothing of the optical flow at the edges of the structure, we assigned a small value to $\omega_{data}$ and $\omega_{smooth}$ to reduce adverse effects. Adding $\mathcal{L}_{data}$ and $\mathcal{L}_{smooth}$ to the loss function as supplementary constraints improves the model's ability to predict optical flow in intensity-varying and texture-free regions, while also reducing the optimization difficulty and the adverse effects of erroneous optical flow.

$\mathcal{L}_{epe}$ was formulated as follows:
\begin{equation}\label{eq4}
  \mathcal{L}_{epe} = \frac{1}{HW}\sum{\left\|f_{gt}-f_{pre}\right\|}_1,
\end{equation}
where $\frac{1}{HW}\sum$ refers to the average of all elements in the matrix and ${\left\|\cdot\right\|}_1$ denotes the L1 norm. Following the brightness constancy and gradient invariance assumptions (\citealt{Brox+Malik+2010}) $\mathcal{L}_{data}$ contains both photometric loss $\mathcal{L}_{photo}$ and gradient loss $\mathcal{L}_{gradient}$.
\begin{equation}\label{eq5}
  \mathcal{L}_{data} = \omega_{photo}\mathcal{L}_{photo} + \omega_{gradient}\mathcal{L}_{gradient}.
\end{equation}
In the implementation, we set $\omega_{photo}$ and $\omega_{gradient}$ to 1. $\mathcal{L}_{photo}$ guides optical flow prediction and aligns blocks of images with similar appearances by penalizing photometric dissimilarity, which was defined as:
\begin{equation}\label{eq6}
  \mathcal{L}_{photo} = \frac{1}{HW}\sum{\psi \left( I_2-\varphi\left( I_1,f_{pre}\right) \right)},
\end{equation}
where $\psi\left(x\right)={\left( \left| x\right| + \epsilon \right)}^q$ is a robust loss with $\epsilon=0.01$ and $q=0.4$ in the implementation (see \citealt{Liu+etal+2019}) and $\varphi\left( I_1,f_{pre}\right)$ denotes that $I_1$ is warped toward $I_2$ using $f_{pre}$. The matching relationship cannot be accurately calculated by relying on the photometric constancy assumption alone because of intensity variations; therefore, $\mathcal{L}_{data}$ is supplemented by $\mathcal{L}_{gradient}$, which is defined as:
\begin{equation}\label{eq7}
  \mathcal{L}_{gradient} = \frac{1}{HW}\sum{\psi \left(\partial I_2-\partial\varphi\left( I_1,f_{pre}\right) \right)},
\end{equation}
where $\partial I$ denotes the gradient matrix of I. For solar H$\alpha$ or TiO images captured by NVST, $\mathcal{L}_{gradient}$ alleviates the problem that intensity values are not always reliable, making the method robust against intensity variations (\citealt{Brox+etal+2004}).

The above loss function does not take into account the pixel-to-pixel interactions. Therefore, $\mathcal{L}_{smooth}$ is introduced to constrain the smoothness of the optical flow field, which is formulated as follows:
\begin{equation}\label{eq8}
  \mathcal{L}_{smooth} = \frac{1}{HW}
  \left( 
  \sum{{\left\| \partial f_{pre} \right\|}_1 e^{-\left| \partial I_2 \right|}} + 
  \sum{{\left\| \partial^2 f_{pre} \right\|}_1 e^{-\left| \partial^2 I_2 \right|}}
  \right).
\end{equation}
$\mathcal{L}_{smooth}$ constrains the interactions between neighboring pixels, which enhances the similarity of optical flow in neighboring regions, making the optical flow field more realistic and producing more accurate optical flow in regions without texture.

RAFT has revealed that by increasing the output of the intermediate optical flow and incorporating these outputs into the calculation of loss, the difficulty of optimizing a single high-resolution output can be reduced and the accuracy can be improved (\citealt{Teed+Deng+2020}). PWCNet uses multi-scale outputs and assigns different weights to the outputs at different scales when calculating the total loss. However, it uses small weight factors for high-resolution prediction losses and larger weight factors for low-resolution prediction losses. \citet{Brox+Malik+2010} demonstrated that assigning decreasing weights to secondary factors allows training to converge quickly in the early stages without affecting the refinement in the later stages. This idea can be applied to the training loss of multi-scale output. We assigned small weight factors to the coarse-scale prediction losses and large weight factors to the fine-scale prediction losses, enabling the network to rapidly converge at the coarse scale and provide good initialization for the subsequent refinement of the optical flow. The above training scheme accelerates the convergence speed of the network, enhances the prediction ability of the network for fine-scale optical flow, and achieves accurate prediction of small-target optical flow.

For the entire prediction loss sequence 
$\left\{ \mathcal{L}_1,\mathcal{L}_2,\mathcal{L}_3,\mathcal{L}_4 \right\}$, 
we assign decreasing weights, and the total training loss $\mathcal{L}$ is formulated as follows:
\begin{equation}\label{eq9}
  \mathcal{L} = \sum_{i=1}^{4}{\gamma^{i-1}\mathcal{L}_i},
\end{equation}
where $\gamma^{i-1}$ is the weight factor. In the experiment, we set $\gamma$ to 0.5,

\section{EXPERIMENTS AND ANALYSIS}
\label{sect:exp}

\subsection{Experimental Environment}
The hardware used in the experiment was an AMD EPYC 7543 32-Core Processor CPU @ 3.7 GHz with a NVIDIA RTX A5000 graphics card, 30 GB of memory, Ubuntu20.04LTS operating system, and PyCharm. Python was used as the programming language, and the software environment was based on CUDA11.0, Cudnn11.0, and Pytorch1.7.1 deep-learning frameworks.

\subsection{Evaluation Metrics}
In this study, we evaluated the performance of the model by using the average-endpoint error (AEPE) on the proxy dataset, which was generated from Dataset 1 by applying Demons. The AEPE measures the consistency of the model-estimated optical flow with the proxy optical flow, with smaller values indicating better performance.
\begin{equation}\label{eq10}
  AEPE = \frac{1}{N} \times \sum_{i=1}^{N}{\frac{1}{HW} 
  \sum{ {\left\| f_{pre}-f_{gt} \right\|}_2 } },
\end{equation}
where N denotes the number of input images.

For Datasets 1 and 2, since the true optical flow is not available, we first used the predicted optical flow between two consecutive frames to generate the predicted image $\hat{I}_1$ through the function $\varphi\left( I_1,f_{pre}\right)$. Then, we employed the evaluation metrics of structural similarity (SSIM, \citealt{Wang+etal+2004}), correlation coefficient (CC), residual variance (RV), and residual mean (RM) to assess the alignment quality of $\hat{I}_1$ and evaluate the accuracy of the predicted optical flow. The formula for the CC calculation is as follows:
\begin{equation}\label{eq11}
  CC = \frac{\sum{\left(I_1 - \overline{I}_1 \right)} \left(I_2 - \overline{I}_2 \right)}
  {\sqrt{\sum{\left(I_1 - \overline{I}_1 \right)}^2 \sum{\left(I_2 - \overline{I}_2 \right)}^2}}, 
\end{equation}
where $\overline{I}$ is the mean value of $I$.

The larger the values of SSIM and CC, the more similar the two images are in structure and the more accurate the optical flow. On the other hand, the smaller the values of RV and RM, the closer the appearance of the two images, and the more robust the model performance.

\subsection{Experimental Results and Analysis}
During the training process, the AdamW optimizer (\citealt{Loshchilov+Hutter+2017}) was utilized and the gradients were clipped within the range $\left[ -1,1 \right]$. The maximum learning rate was set to 0.001 and the learning rate weight decay was set to 0.0001. The batch size was set to 2 and the learning rate adjustment scheme followed OneCycleLR (\citealt{Smith+Topin+2019}). Data augmentation techniques followed that of RAFT, including random flip, rotation, photometric augmentation, stretching, and other data augmentation methods. These augmentation techniques were used to improve the model's generalization performance and increase its robustness to illumination changes.

In this study, we performed a set of comparison and ablation experiments to demonstrate that the proposed method effectively solves the optical flow prediction issue for small targets in high-resolution solar images. All comparison methods were trained using the same proxy dataset, and the training scheme was consistent for all models to measure the performance of each model. Tables 1 and 2 show the comparison of AEPE, SSIM, CC, RV, and RM on Datasets 1 and 2.

Our method outperforms all previous methods as shown in Table~\ref{Tab1}. We reduced the AEPE value from 1.13 of RAFT to 0.52, showing a 54\% improvement. The SSIM also improved by 4.4\% from 0.9182 in Demons to 0.9587. Our method achieved a CC of 0.9930, RV of 15.2818, and RM of 3.1508, which are all better than the results from previous methods.

\begin{table}
\begin{center}
\caption[]{ Performance of Different Optical Flow Methods on Dataset 1.}
\label{Tab1}


 \begin{tabular}{lccccc}
  \hline\hline\noalign{\smallskip}
Method                                       & AEPE  & SSIM      & CC        & RV        & RM        \\
  \hline\noalign{\smallskip}
Demons (\citealt{Liu+etal+2018})             & -     & 0.9182    & 0.9870    & 24.9931   & 4.4918    \\
FlowNet (\citealt{Dosovitskiy+etal+2015})    & 1.99  & 0.8433    & 0.9666    & 53.6588   & 7.9483    \\
PWCNet (\citealt{Sun+etal+2018})             & 1.26  & 0.9042    & 0.9789    & 33.1684   & 6.2046    \\
RAFT (\citealt{Teed+Deng+2020})              & 1.13  & 0.8666    & 0.9716    & 47.3137   & 7.1653    \\
Ours                                        & $\mathbf{0.52}$  & $\mathbf{0.9587}$    & $\mathbf{0.9930}$    & $\mathbf{15.2818}$   & $\mathbf{3.1508}$    \\
  \noalign{\smallskip}\hline
\end{tabular}
\end{center}
\end{table}

It can be observed from Fig.~\ref{Fig1} that there is a significant difference in appearance between Datasets 1 and 2, and Dataset 2 is not involved in model training. Despite this, as seen in Table~\ref{Tab2}, our method still shows efficacy in predicting optical flow in Dataset 2, indicating the network's ability to extract crucial features from solar images beyond just appearance. Our method outperforms others in all metrics, especially SSIM, RV, and RM, achieving 3.6\%, 34.5\% and 33.7\% improvement respectively compared to the suboptimal method (PWCNet). These results demonstrate that our framework has strong generalization and accurate prediction abilities for small-target optical flows in solar images.

Notably, the AEPE of our method in Tables 1 and 2 are not zero, while SSIM and other indexes are significantly better than Demons, indicating that: 1) there are errors in the optical flow predicted by Demons; 2) although the optical flow data generated by Demons are used as the proxy data for training, the accuracy of optical flow predictions is corrected because of the use of multi-channel outputs and the unsupervised loss strategy in the model.

\begin{table}
\begin{center}
\caption[]{ Performance of Different Optical Flow Methods on Dataset 2.}
\label{Tab2}
 \begin{tabular}{lccccc}
  \hline\hline\noalign{\smallskip}
Method                                      & AEPE  & SSIM      & CC        & RV        & RM        \\
  \hline\noalign{\smallskip}
Demons (\citealt{Liu+etal+2018})             & -     & 0.9084    & 0.9763    & 33.9556   & 6.0755    \\
FlowNet (\citealt{Dosovitskiy+etal+2015})    & 0.72  & 0.8815    & 0.9600    & 51.9735   & 8.3302    \\
PWCNet (\citealt{Sun+etal+2018})             & 0.57  & 0.9289    & 0.9781    & 25.9908   & 5.8943    \\
RAFT (\citealt{Teed+Deng+2020})              & 0.56  & 0.8896    & 0.9636    & 47.2199   & 7.9377    \\
Ours                                         & $\mathbf{0.24}$  & $\mathbf{0.9628}$    & $\mathbf{0.9893}$    & $\mathbf{17.0194}$   & $\mathbf{3.9101}$    \\
  \noalign{\smallskip}\hline
\end{tabular}
\end{center}
\end{table}

Our proposed method improves upon PWCNet by making the following improvements: (1) adjusting the number of feature layers (NFL), (2) adding residual blocks (RB), (3) using multi-output modules (MLO), and (4) improving loss functions (ILF). To evaluate the impact of each improvement, we conducted an ablation study on the test set of Dataset 1. Results are shown in Table~\ref{Tab3}, which highlights the contribution of each module and confirms the effectiveness of the proposed method.

\begin{table}
\begin{center}
\caption[]{ Performance of Different Improvements on Dataset 1.}
\label{Tab3}
 \begin{tabular}{cccc|cccccc}
  \hline\hline\noalign{\smallskip}
NFL         & RB        & MLO       & ILF        & AEPE  & SSIM      & CC        & RV        & RM        & Params\\
  \hline\noalign{\smallskip}
            &           &           &           & 1.27  & 0.8776    & 0.9735    & 42.0205   & 6.9835    & 5.4M\\
$\surd$     &           &           &           & 0.51  & 0.9469    & 0.9901    & 18.8928   & 4.0131    & 9.8M\\
$\surd$     & $\surd$   &           &           & 0.54  & 0.9519    & 0.9914    & 16.9177   & 3.6996    & 4.2M\\
$\surd$     & $\surd$   & $\surd$   &           & 0.51  & 0.9522    & 0.9915    & 16.6356   & 3.6873    & 5.1M\\
$\surd$     & $\surd$   & $\surd$   & $\surd$   & 0.52  & 0.9587    & 0.9930    & 15.2818   & 3.1508    & 5.1M\\
  \noalign{\smallskip}\hline
\end{tabular}
\end{center}
\end{table}

As can be seen in Table~\ref{Tab3}, using NFL significantly improved the results of all metrics compared to PWCNet. The proxy dataset and Dataset 1 are high-resolution H$\alpha$ images of solar radiation captured by NVST, containing many small targets. Adding full-resolution features to the NFL improved the AEPE value from 1.27 to 0.51 and the SSIM value from 0.8776 to 0.9469. The above results demonstrate that low-level features in high-resolution features contribute to the optical flow prediction of small targets, and the NFL module enhances the accuracy of optical flow prediction.

Adding RB improved all metrics except for AEPE. The shortcut connections in the RB helped to retain low-level structural features and improved the model's ability to predict optical flow for small targets. However, the presence of erroneous optical flow in the proxy dataset caused some pixels to be occluded, which increased the AEPE value. Despite this, using RB reduced the number of network parameters and accelerated the convergence of the model.

The use of MLO improved all metrics, reducing the RV value from 16.9177 in the RB to 16.6356. The inception structure in the MLO provided the model with multiple matching relationships that contained different features. Dilated convolution provided the network with sparse information within a larger receptive field, and the network obtained secondary matching relations far from the sampling point. The results show that MLO reduced the error rate of the network in predicting the optical flow with a slight increase in the parameters.

Finally, ILF improved all metrics except for AEPE. On the one hand, $\mathcal{L}_{photo}$ in ILF made the predicted image $\hat{I}_1$ and target image $I_2$ as visually similar as possible, while $\mathcal{L}_{smooth}$ enhanced the connection of adjacent optical flows, which conforms to reality. By ILF, the method achieved an SSIM of 0.9587, RV of 15.2818, and RM of 3.1508, all of which are better than those obtained in the other configurations. The results demonstrate that ILF reduces the negative impact of erroneous optical flow in the proxy dataset and improves the overall performance of the model. This approach allows for training the network with proxy optical flow labels, which are easier to obtain than truth optical flow.

\begin{table}[H]
\begin{center}
\caption[]{ Performance of Demons and Our Method on a Local Area of the Frame in Dataset 1.}
\label{Tab4}


 \begin{tabular}{lcccc}
  \hline\hline\noalign{\smallskip}
Method                                      & SSIM      & CC        & RV        & RM        \\
  \hline\noalign{\smallskip}
Demons (\citealt{Liu+etal+2018})             & 0.8570    & 0.9824    & 50.7849   & 6.7924    \\
Ours                                        & 0.9187    & 0.9891    & 33.4745   & 5.2279    \\
  \noalign{\smallskip}\hline
\end{tabular}
\end{center}
\end{table}

   \begin{figure}[htbp]
   \centering
	\begin{subfigure}{0.23\linewidth}
		\includegraphics[width=1\linewidth]{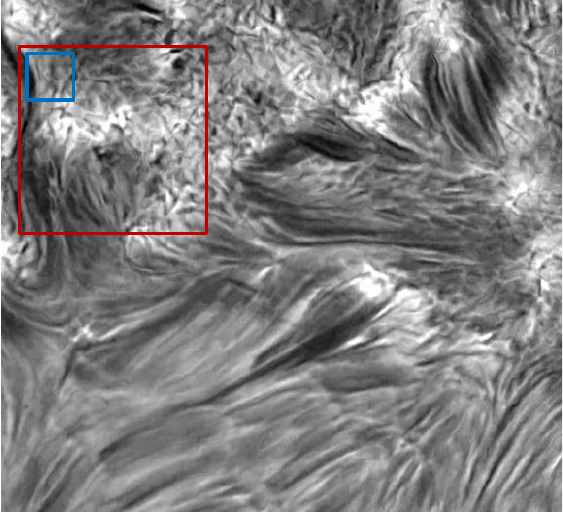}
		\caption{Demons}
		\label{warp-a}
	\end{subfigure}
    \hspace{0.05cm}
	\begin{subfigure}{0.23\linewidth}
		\centering
		\includegraphics[width=1\linewidth]{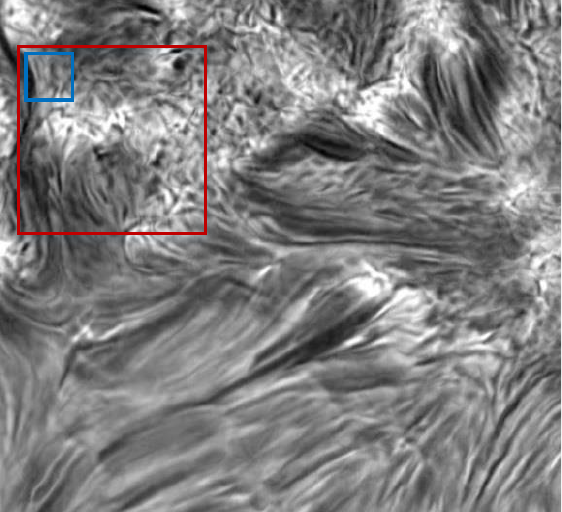}
		\caption{PWCNet}
		\label{warp-b}
	\end{subfigure}
    \hspace{0.05cm}
	\begin{subfigure}{0.23\linewidth}
		\centering
		\includegraphics[width=1\linewidth]{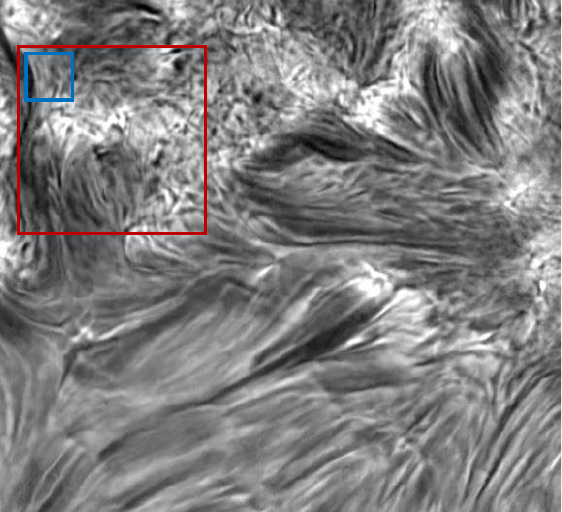}
		\caption{RAFT}
		\label{warp-c}
	\end{subfigure}
	\hspace{0.05cm}
    \begin{subfigure}{0.23\linewidth}
	    \centering
	    \includegraphics[width=1\linewidth]{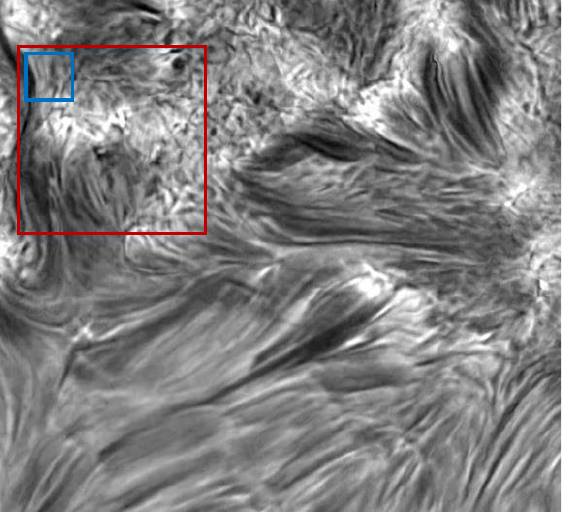}
	    \caption{Ours}
	    \label{warp-d}
    \end{subfigure}
    \vspace{-0.3cm}
	\caption{Predicted image for different methods.}
	\label{Fig7}
   \end{figure}

    \begin{figure}[htbp]
   	\centering
   	\begin{subfigure}{0.23\linewidth}
   		\centering
   		\includegraphics[width=1\linewidth]{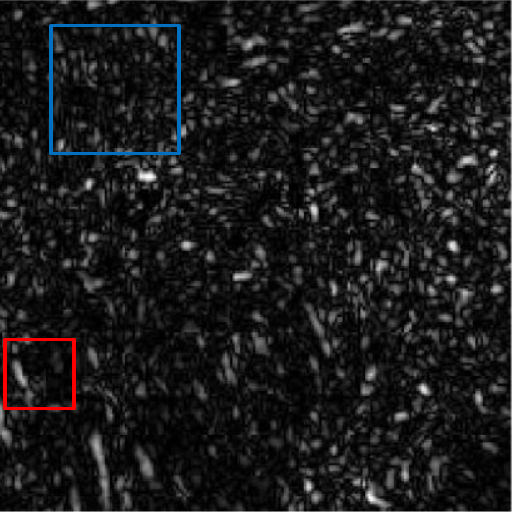}
   		\caption{Demons}
   		\label{Res-a}
   	\end{subfigure}
	\hspace{0.05cm}
   	\begin{subfigure}{0.23\linewidth}
   		\centering
   		\includegraphics[width=1\linewidth]{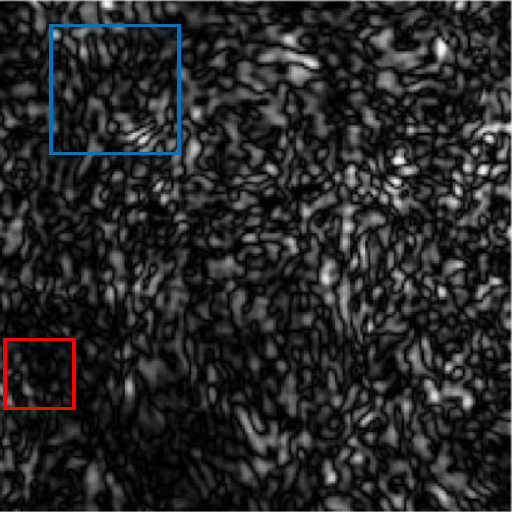}
   		\caption{PWCNet}
   		\label{Res-b}
   	\end{subfigure}
	\hspace{0.05cm}
   	\begin{subfigure}{0.23\linewidth}
   		\centering
   		\includegraphics[width=1\linewidth]{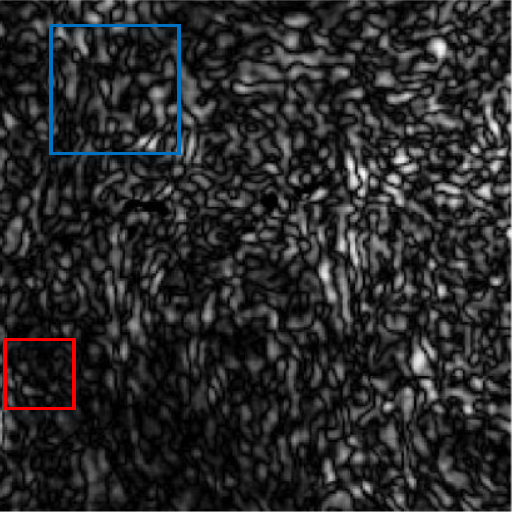}
   		\caption{RAFT}
   		\label{Res-c}
   	\end{subfigure}
    \hspace{0.05cm}
   	\begin{subfigure}{0.23\linewidth}
   		\centering
   		\includegraphics[width=1\linewidth]{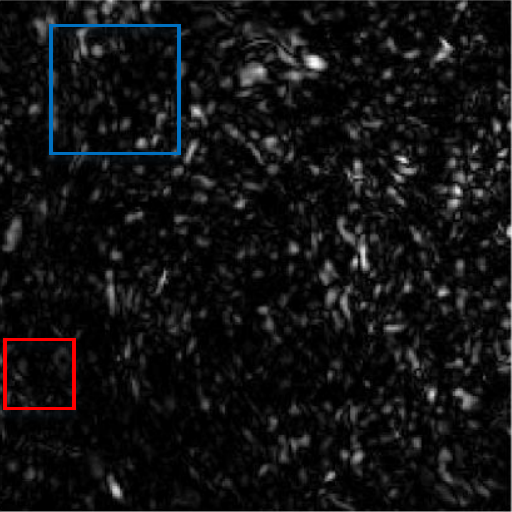}
   		\caption{Ours}
   		\label{Res-d}
   	\end{subfigure}
   	\vspace{-0.3cm}
   	\caption{Residual map for different methods.}
   	\label{Fig8}
   \end{figure}

   \begin{figure}[htbp]
   	\centering
   	\begin{subfigure}{0.23\linewidth}
   		\centering
   		\includegraphics[width=1\linewidth]{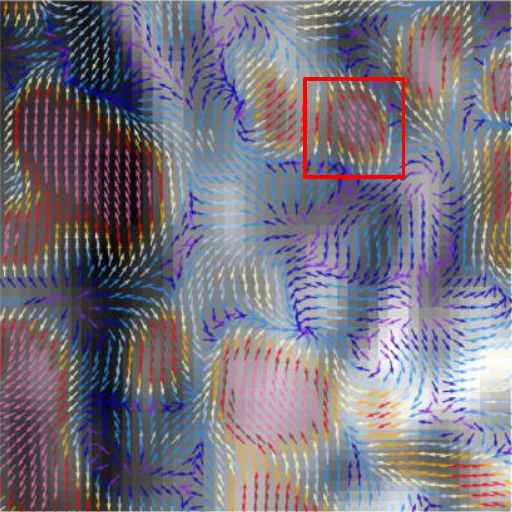}
   		\caption{Demons}
   		\label{Arrow-a}
   	\end{subfigure}
   	\hspace{0.05cm}
   	\begin{subfigure}{0.23\linewidth}
   		\centering
   		\includegraphics[width=1\linewidth]{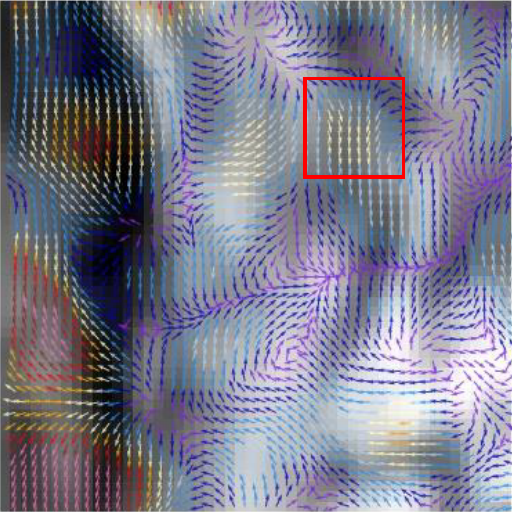}
   		\caption{PWCNet}
   		\label{Arrow-b}
   	\end{subfigure}
   	\hspace{0.05cm}
   	\begin{subfigure}{0.23\linewidth}
   		\centering
   		\includegraphics[width=1\linewidth]{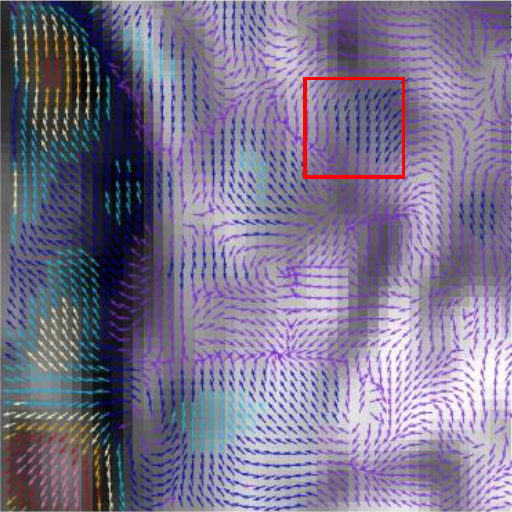}
   		\caption{RAFT}
   		\label{Arrow-c}
   	\end{subfigure}
    \hspace{0.05cm}
   	\begin{subfigure}{0.23\linewidth}
   		\centering
   		\includegraphics[width=1\linewidth]{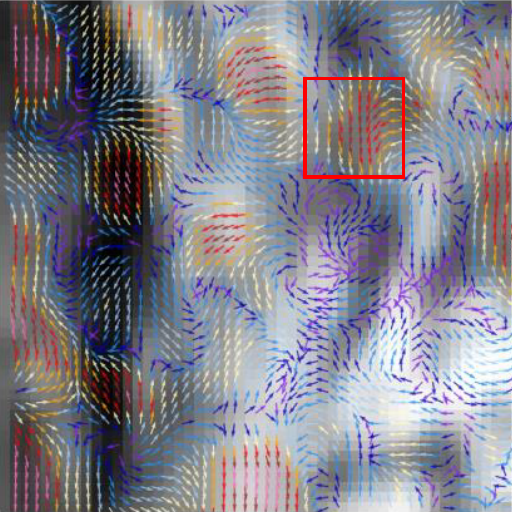}
   		\caption{Ours}
   		\label{Arrow-d}
   	\end{subfigure}
   	\vspace{-0.3cm}
   	\caption{Arrow map for different methods.}
   	\label{Fig9}
   \end{figure}

   \begin{figure}[htbp]
   	\centering
   	\begin{subfigure}{0.24\linewidth}
   		\centering
   		\includegraphics[width=0.99\linewidth]{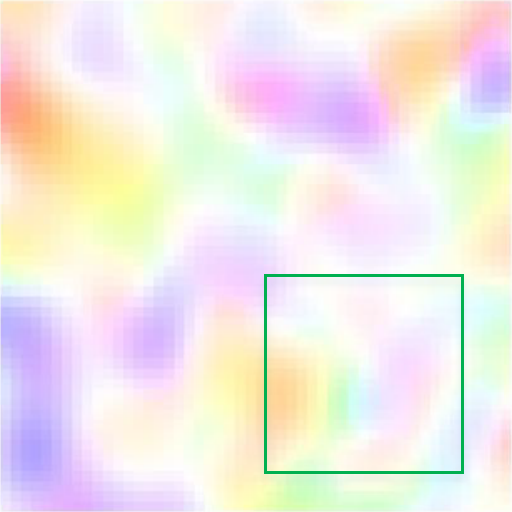}
   		\caption{Demons}
   		\label{Flow-a}
   	\end{subfigure}
   	\centering
   	\begin{subfigure}{0.24\linewidth}
   		\centering
   		\includegraphics[width=0.99\linewidth]{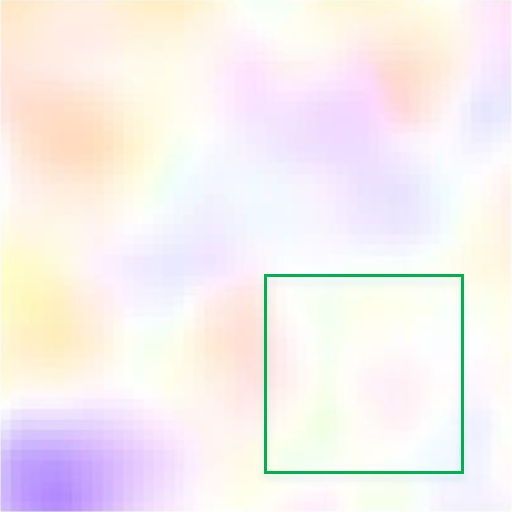}
   		\caption{PWCNet}
   		\label{Flow-b}
   	\end{subfigure}
   	\centering
   	\begin{subfigure}{0.24\linewidth}
   		\centering
   		\includegraphics[width=0.99\linewidth]{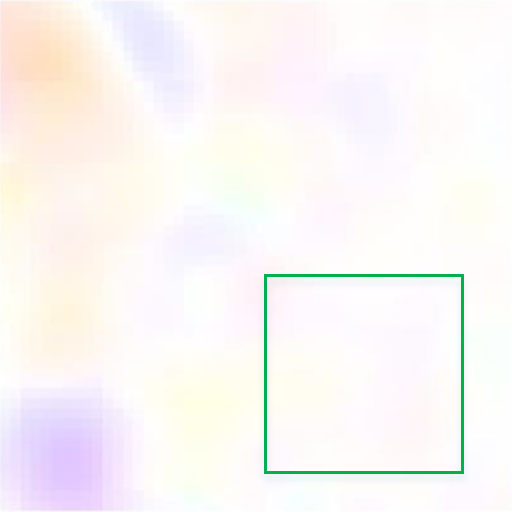}
   		\caption{RAFT}
   		\label{Flow-c}
   	\end{subfigure}
   	\begin{subfigure}{0.24\linewidth}
   		\centering
   		\includegraphics[width=0.99\linewidth]{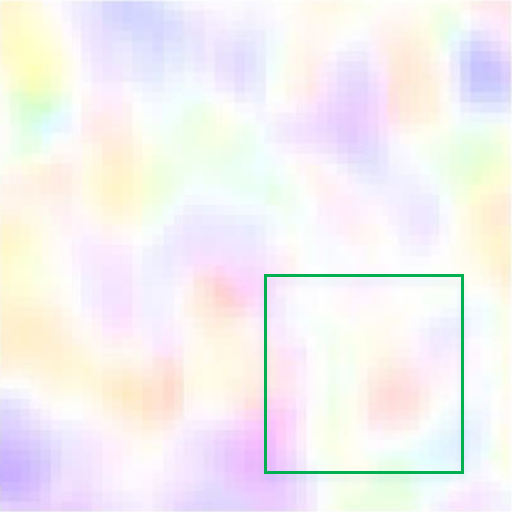}
   		\caption{Ours}
   		\label{Flow-d}
   	\end{subfigure}
   	\vspace{-0.3cm}
   	\caption{Optical flow visual map for different methods.}
   	\label{Fig10}
   \end{figure}

Fig.~\ref{Fig7}--\ref{Fig10} show visual comparisons of different methods on Dataset 1, including machine-learning methods such as Demons, and deep-learning methods such as PWCNet and RAFT. Fig.~\ref{Fig7} displays the predicted images generated by each method for Dataset 1. Fig.~\ref{Fig8} shows the residuals of the target and predicted images in the red boxed area with a $200\times200$ pixel size in Fig.~\ref{Fig7}. Fig.~\ref{Fig8} shows that the proposed method and Demons eliminate a large number of fragmented grayscale differences, demonstrating the capability of the proposed method in predicting tiny target optical flow and outperforming PWCNet and RAFT. To further compare the performance of the proposed method and Demons, Table~\ref{Tab4} provides a quantitative comparison of the two methods for optical flow estimation in the red box region of Fig.~\ref{Fig7}. The results in Table~\ref{Tab4} show that our method outperformed Demons in all metrics, indicating that our method is better at predicting the optical flow of high-resolution solar images.

Fig.~\ref{Fig9} shows a vector arrow plot of the estimated optical flow for the blue boxed region in Fig.~\ref{Fig8}, with the color of the arrows indicating the magnitude of the optical flow. In the region highlighted with red boxes in Fig.9, the optical flow predicted by PWCNet and RAFT had no significant difference, whereas the optical flow predicted by Demons and our method had significantly different and smaller residuals. This suggests that PWCNet and RAFT struggle to accurately capture the motion of small, pixel-level structures and predict their optical flow, while Demons and the proposed method perform better in this regard. Fig.~\ref{Fig10} visualizes the optical flow for the blue boxed region in Fig.8, with color indicating the magnitude and direction of the flow. The green boxed region in Fig.~\ref{Fig10} shows that PWCNet and RAFT generate blurred optical flow boundaries, whereas Demons and our method produce clearer boundaries, further highlighting the superiority of our proposed method in accurately predicting the optical flow of small targets.

\section{CONCLUSIONS}
\label{sect:Cons}
The measurement of the transverse velocity field of high-resolution solar images plays a crucial role in understanding the dynamic characteristics of the solar magnetic field and atmosphere and predicting solar activity. To address the problem of the low accuracy of optical flow prediction in high-resolution solar images, we proposed an improved PWCNet model based on CNN. Our method adopted NFL, RB, MLO, and ILF to enhance the ability of the model to predict the optical flow of small targets and improve the accuracy of the algorithm.

Experimental results showed that our proposed method was highly effective in predicting the optical flow of small targets on solar H$\alpha$ and TiO images captured by the NVST telescope. The accuracy of our optical flow prediction outperformed that of other main optical flow methods. Our proposed method has potential applications for velocity field measurements of solar images captured by other astronomical telescopes. However, it is worth noting that the optical flow prediction accuracy in textureless regions is not high and may be due to errors in the proxy dataset or limitations in our framework. Further research is required to resolve these issues.

\label{lastpage}

\begin{acknowledgements}
This study is supported by the National Natural Science Foundation of China under Grant Nos. 12063002, 12163004, 12073077.
\end{acknowledgements}

\bibliographystyle{raa}
\bibliography{bibtex}

\end{document}